\documentstyle[prb,aps,twocolumn]{revtex}
\begin{document}
\draft
\title
{Ultrafast optical signature of quantum superpositions in a
nanostructure}
\author{F.J.Rodr\'{\i}guez$^{1}$, L.Quiroga$^{1}$
and N.F.Johnson$^{2}$}
\vspace{.05in}
%
\address
{$^{1}$ Departamento de F\'{\i}sica, Universidad de Los Andes, A.A.
4976,
Bogot\'a D.C., Colombia\\
$^{2}$ Physics Department, Clarendon Laboratory,
Oxford University, Oxford OX1 3PU, U.K.}
%
\maketitle

\begin{abstract}
We propose an unambiguous signature for detecting quantum
superposition states in a nanostructure, based on current
ultrafast spectroscopy techniques. The reliable generation
of such superposition states via Hadamard-like quantum gates is crucial for
implementing solid-state based quantum information schemes.  The signature
originates from a remarkably strong photon antibunching effect which is
enhanced
by non-Markovian dynamics.
\end{abstract}


\narrowtext
The possibility of performing quantum information processing in
nanostructure
systems is of great interest from the perspectives of
both fundamental science and future emerging technologies.
Quantum dots (QDs) are solid-state nanostructures which are
analogs of real atoms\cite{hawrylak}; semiconductor QDs and macromolecules are seen as
excellent candidates for performing quantum information processing
tasks\cite{quiroga,imamoglu,loss,biolatti}. There are also many inorganic and
organic structures that qualify as `nanostructures' and may therefore
become good candidates: e.g. carbon buckyballs, and even
micro-biological
molecular structures such as the photosynthetic complexes in purple
bacteria\cite{schulten}.  Essential steps toward the implementation of
standard
quantum information schemes in such nanostructures, include: (1) the
identification of the basic qubit (quantum bit), and (2) the application of
one-
and two-qubit quantum gates in order to generate quantum superpositions and
entanglement. An important example of a one-qubit gate is the Hadamard-like
gate, since it generates a superposition state (e.g. $|0\rangle+|1\rangle$)
from an un-superposed initial state (e.g. $|0\rangle$). Reference [2] 
showed that (1) and (2) can be achieved with excitons
generated using current techniques in ultrafast optical
spectroscopy\cite{bonadeo}. Several experimental groups are now
actively pursuing this route. Such qubit control has already been achieved
in neutral atoms\cite{atoms} and ions in an ion trap\cite{cirac}, however scalability
issues may limit such non solid-state devices to just a few qbits. 

However there is a crucial third step, which is to verify the reliability of the
quantum superpositions (entanglements) generated in the single nanostructure
(pair of nanostructures) by the one-qubit (two-qubit) quantum gate.  
How can we show experimentally that we have generated such
superposition
(entanglement)? The present work addresses this question for the important first
step of a single nanostructure (e.g. a single QD).  Specifically, we
consider
the ultrafast second-order coherence function of the emitted light from the
optically-generated exciton in a single nanostructure (e.g. a QD). This
quantity 
$g^{(2)}$ is calculated for the QD interacting with two baths: (i)
photon environment and (ii) a phonon system. A
strong antibunching effect is predicted in the resonance fluorescence
response
at very
short times, {\it if and only if the initial exciton state comprises a
quantum superposition}. This strong effect does not arise for initial
states described by a statistical mixture, i.e. non-superposed states.
We show numerically, that non-Markovian effects
significantly enhance the antibunching signal, hence demonstrating
that temporal correlations {\em cannot} be neglected a priori in such ultrafast regimes. 
In addition to the quantum
superposition test, our results may prove useful in designing photon-emitting devices with
controllable and accurate emission rates\cite{yamamoto}.

In atomic systems, resonance fluorescence experiments have
already proved themselves to be extremely valuable\cite{scully}.
Similar experiments in solid-state systems have only recently been
performed. In particular it has been demonstrated that for a single CdSe QD
at
room temperature\cite{imamoglu1,cds} and a single self-assembled
InAs QD at cryogenic temperatures\cite{imamoglu2,gershoni}, strong antibunching
effects
are observed in fluoresence experiments. This provides direct evidence
that single QDs present the same kind of nonclassical light emission as a
single two-level atom.  The present paper provides a further stimulus to 
experimentalists to improve resolution times, with the potential payoff that
antibunching measurements will then provide a direct probe of the initial
quantum state.

The solid-state system of interest here comprises a
nanostructure (e.g. a QD) of any shape,
coupled to the electromagnetic field and to a heat bath, represented by a
set of
harmonic oscillators, which provides the basic source of temperature
dependence. 
The light source is of low intensity 
hence the number of excitons generated is small. 
A single exciton in its ground state can be described by a two-level
system\cite{quiroga}. The Hamiltonian is
given by
\begin{eqnarray}
H=\frac{\epsilon}{2} \sigma_z+\sum_k{\omega_k a_k^{\dag}a_k}+
\sum_k{(g_k\sigma^+a_k+g_k^*\sigma^-a_k^{\dag})}\\ \nonumber
+\sum_q{\Omega_qb_q^{\dag}b_q}+
\sum_q{\sigma_z(f_qb_q+f_q^*b_q^{\dag})}+\\ \nonumber
E(t)\sigma^+e^{i\omega t}+c.c.
\end{eqnarray}
where $a_k$ and $b_q$ represent the annihilation operators for photons
and phonons respectively, $\epsilon$ is the total exciton energy,
$\omega_k$ ($\Omega_q$) denotes the photon (phonon) frequencies,
$g_k$ ($f_q$) the exciton-photon (exciton-phonon) coupling and $E(t)$
describes
the envelope of
a classical source of light of frequency $\omega$ acting on the QD.
The
exciton population is described by $\sigma_z=|X\rangle \langle
X|-|0\rangle\langle 0|$ where $|X\rangle$ stands for a one exciton state
while
$|0\rangle$ denotes the QD vacuum, i.e. no exciton. Similarly
the raising and lowering pseudo-spin operators are
$\sigma^+=|X\rangle \langle 0|$ and $\sigma^-=|0\rangle \langle X|$,
respectively.
From Eq.(1) it is seen
that the photon field is associated with the dissipative dynamics of the
QD
whereas the phonon field is responsible for dephasing effects.
This phonon dephasing accounts for the
temperature effects. The fundamental band gap is typically much larger than $k_B T_e$
($T_e$ is the temperature) hence the photon field can be taken as remaining
at zero temperature. Non-Markovian effects are included for both exciton-photon and
exciton-phonon couplings. One of the advantages of this model is
that it is reasonably simple, yet sufficiently complex to
manifest many important features of the ultrafast response of
nanostructures.  More refined models should consider multiexciton complexes 
if a strong excitation is
applied.  In that case, joint effects coming from Non-Markovian processes as
described here, together with strong  particle correlations, should be observable.
However, these kind of effects are beyond of the scope of the present work.

We employ a master equation of Lindblad form, since 
this kind of equation can properly account for the coupling
of the QD system to its environment - it can also go beyond the Markov
approximation using time dependent damping
coefficients\cite{petruccione}.
We stress that master equations with time-independent damping coefficients are unable to account
for the evolution of an open system on very short time scales.
At resonance, i.e. $\omega=\epsilon$, and using the rotating wave
approximation, the Liouvillian acting on any QD operator $O$ is given by:
\begin{eqnarray}
L(t)O=-\frac {i}{\hbar}[E\sigma^++E^*\sigma^-,O]+\\ \nonumber
\gamma_{relax}(t)[\sigma^-O\sigma^+-
\frac {1}{2}O\sigma^+\sigma^--\frac {1}{2}\sigma^+\sigma^-O]\\ \nonumber
+\gamma_{dph}(T_e,t)[\sigma_z,[\sigma_z,O]]
\end{eqnarray}
where the coupling to photons $\gamma_{relax}(t)$, and the coupling
to phonons $\gamma_{dph}(T_e,t)$, include
non-Markovian effects through their time dependences.  
Temperature effects do not need to be included for $\gamma_{relax}$ in the coupled light-QD subsystem, since $\hbar
\epsilon >> k_B T_e$. Solving the Liouville equation for the QD density matrix $\rho$, and for different
Rabi frequencies $\Omega=\frac{\mu|E|}{\hbar}$ ($\mu$ is the dipole moment),
expectation values for any QD operator
may be evaluated and the characteristics of the emitted
photon field hence obtained.

The coherence properties of the emitted photon field can be properly
accounted for by the second-order coherence function given
by\cite{charmichael}
\begin{eqnarray}
g^{(2)}(T,\tau)=\frac {\langle
\sigma^+(T)\sigma^+(T+\tau)\sigma^-(T+\tau)\sigma^-(T)
\rangle}{\langle\sigma^+(T)\sigma^-(T)\rangle \langle
\sigma^+(T+\tau)\sigma^-(T+\tau)
\rangle}\ \ .
\end{eqnarray}
where $\tau=t_1-t_2$ represents the time difference between the photons' arrival at the detector and
$T=\frac{t_1+t_2}{2}$. This coherence function can be expressed in a very simple form, valid for any initial condition
of the QD, as follows:
\begin{eqnarray}
g^{(2)}(T,\tau)=\frac {\langle X|{\cal T}e^{\int_0^{\tau}
L(t^{\prime})dt^{\prime}}(|0\rangle \langle 0|)|X\rangle}
{\rho_{X,X}(T+\tau)}
\end{eqnarray}
where ${\cal T}$ denotes the time-ordering operator and $\rho$ is the QD
reduced
density matrix.
By changing the initial preparation state of the QD, the value of the non-stationary second-order coherence
function
should change through its dependence on $\rho_{X,X}$.
A closed expression for $g^{(2)}$ and the antibunching effect, characterized by the growth of $g^{(2)}$ 
from zero for $\tau=0$, has been well documented {\it in the steady-state} situation, with non-zero relaxation 
decay $\gamma_0$ and with $\gamma_{dph}=0$\cite{charmichael}.
However an expression for Eq.(4) in closed form for the ultrafast regime,
valid for any $T$ and $\tau$, is not available. A numerical solution
of the Bloch equations for the different elements of the QD reduced density
matrix must therefore be performed.
 Our present work represents a far more general study of the variations of $g^{(2)}$, and includes the effect of 
different initial QD
states. In particular,
it is interesting to quantify the variations of $g^{(2)}$ for the following
initial conditions: (i) a pure state comprising a quantum superposition (QS)
of type $|\Psi\rangle=\frac {1}{\sqrt{2}}
(|0\rangle+i|X\rangle)$ where $\rho(0)=|\Psi\rangle \langle \Psi|$,
(ii) the usual experimental case in which the QD is prepared in its ground state (GS)  
given by $\rho(0)=|0\rangle \langle 0|$, and
(iii) a statistical mixture (SM) of states $\rho(0)=\frac{1}{\sqrt{2}}(|0\rangle \langle 0| + |X\rangle \langle X|)$. 
 The initial photon state is the vacuum.

We start by examining the system's qualitative behavior within the Markov approximation. 
Results for $g^{(2)}$ are shown in Fig. 1a for a single QD
containing up to one electron-hole pair. The experimentally obtained decay rate\cite{bonadeo,kamada} 
 $\hbar\gamma_{relax}=\hbar\gamma_0 = 20 \mu eV$ and $\gamma_{dph}=0.5\gamma_0$
are used in the calculations. The Rabi 
frequency is 2.25$\gamma_0$ ($\mu \approx$ 15 Debyes). 
 A clear sub-poissonian character is observed at very short times. The
enhancement property of $g^{(2)}$  can be readily understood from Eq.(4), in particular
from a vanishing value of the element $\rho_{X,X}$ of the QD density matrix. The second-order time 
correlation function
can be written as 
 $g^{(2)}(T,\tau) = (\rho_{X,X}(\tau)|_{\rho(\tau=0)=|X\rangle\langle X|})/\rho_{X,X}(T,\tau)$,
where the numerator represents the density matrix element given that the QD is in its ground state. By solving the Bloch
equations for $\rho$ at very short times, it can be seen that this enhancement appears for 
$\tau\approx (\Omega-\gamma_0)^{-1}$, i.e. when $\rho_{X,X}(T,\tau)\rightarrow 0$ in agreement with Fig.1a.
We stress that this condition {\it cannot} be obtained if the system is initially prepared in its GS
or in a SM of states (not shown). Previous experiments\cite{imamoglu2} with their limited 
resolution times could not have detected this feature because the correlations always vanish for long detection times.

Since this antibunching behaviour occurs at very short times,  we will now consider the quantitative effects 
of non-Markovian behavior characterized by time-dependent damping rates. 
 In reference $[16]$ an explicit
expression for $\gamma_{relax}(t)$, is given as
\begin{eqnarray}
\gamma_{relax}(t)=\frac {2\gamma_0{\rm sinh}(d t/2)}{(d/\lambda){\rm
cosh}(d t/2)+{\rm sinh}(d t/2)}
\end{eqnarray}
where $d=\sqrt{\lambda^2-2\gamma_0\lambda}$, $\gamma_0$ is the constant Markov
decay rate (time-independent) and $\gamma_0/\lambda$ is the ratio between the photon
reservoir correlation
time and a typical time scale on which the QD exciton changes. For  $\gamma_0/\lambda<<1$, this yields 
$\gamma_{relax}(t)/\gamma_0=1-e^{-\lambda t}$.  This explicit expression for $\gamma_{relax}(t)$, is appropriate 
for a Lorentzian photon reservoir\cite{gisin}. A memoryless photon environment corresponds to
$\lambda$ going to infinity in which case the Markov situation is recovered.
It is worth noting that by slightly changing this form of $\gamma_{relax}(t)$,
non-Markovian effects in a structured photon environment such as a microcavity
could also be modeled\cite{petruccione}.

 For the pure dephasing rate, the standard form of the independent boson model\cite{mahan}
is taken as
\begin{eqnarray}
\gamma_{dph}(T_e,t)=\sum_q{|f_q|^2{\rm coth}(\frac
{\Omega_q}{2T_e})\frac{1-{\rm cos}(\Omega_qt)}{\Omega_q^2}}\ \ .
\end{eqnarray}
In the continuum limit for phonon ${\vec q}$ vectors, all information about
the bath which is essential to
the dynamics of the QD, is contained in the compact form of the spectral
density function
$J(\omega)=\frac {\pi}{2} \sum_q {|f_q|^2\delta(\omega-\Omega_q)}$.
An appropriate choice for the spectral function $J(\omega)$, and its
associated
cut-off frequency, can be made according to the QD
environment. In terms of the Debye model the natural cutoff is the Debye
frequency ($\omega_D$)
yielding a spectral function of the form $J(\omega) \sim \omega^3e^{-\omega/\omega_D}$.
An explicit form for $\gamma_{dph}$, with the latter choice for
$J(\omega)$, has
been derived by Palma et al.\cite{palma}. For different temperatures ($\eta=T_e/\omega_D$),
$\gamma_{relax}(t)$ and $\gamma_{dph}$ are shown in Fig.1b.
On very short time-scales
the effective decay rates for both processes, radiative and pure
dephasing,
are very low indicating that Markov approximations (which are valid
on long time-scales) overestimate the decay effects at short
times.
As demonstrated, this explicit form of $\gamma_{relax}(\tau)$ and $\gamma_{dph}(T_e,\tau)$
leads to an inefficient damping at times $\tau<\gamma_0^{-1}$; hence
the Markov approximation overestimates the damping effects in the ultrafast
regime. Therefore,  any antibunching enhancement that
occurs on a very short time scale, should be reinforced by non-Markovian
phenomena.
A demonstration of this statement is shown in Fig. 1c with $\eta=2.85$ and $\Omega=2.25\gamma_0$, for
different resolution detecting times. Clearly, the QS shows a strong enhancement compared with a GS.
The second-order time correlation function at very short times therefore provides information
about the initial state preparation. The main result of the present work
is that for times smaller than the time scale in which dissipation of
energy takes place, a strong antibunching effect is predicted. Within  this time scale the correlations
between the dot and the environment are not very important and clearly non classical light emission is
enhanced.
As a consequence, antibunching enhancement should be most prominent in QDs
fabricated from II-VI semiconductors, III-V nitrides, organic materials or hybrid
heterostructures
due to the fact that in these systems exciton-photon coupling is not
necessarily weak and non-Markovian effects can be important.

Figure 2 shows how the peak in $g^{(2)}(T\gamma_0=0.5,\tau)$ develops a symmetric shape,
 growing quite large at low Rabi frequencies. For large Rabi frequencies the  antibunching for 
QS initial state is observed at
very short times and the light emission begins to evolve into a mere statistical mixtures of states.
Crucial information concerning the state in which the system was initially prepared, can hence be obtained 
via a measurement of $g^{(2)}$. Our results therefore show that QS states can be far more robust to
decoherence than other initial states at very short times. We conjecture that this robustness may also
hold for multiple qubits (see also Ref.$[22]$).
We now turn our attention to the long time behaviour ($\tau >> \gamma_0^{-1}$) for $g^{(2)}(\tau)$. In the inset
of Fig. 2, the antibunching effect is clearly observed, however the peak intensities are not as large as those
predicted for the QS state. To avoid confusion concerning this sub-Poissonian characteristic, we stress
that this peak intensity comes from the natural Rabi oscillations of the coupled QD-light subsystem.
The antibunching signal reported in this work is observed when the vacuum fluctuations dominate, if and only if the system
is prepared in a QS. Interestingly the QS initial state
is rather insensitive to the decoherence mechanism, for typical experimental values and at very short times.
Experimentally, the effects presented here could be observed using ultrafast spectroscopy with a
resolution of the order of the dephasing times, e.g. 50 ps for a semiconductor QD. We also note that different
initial QS states can be probed by tailoring the polarization of the driving field.

In summary, we have shown that ultrafast fluorescence intensity-correlation
measurements in single QDs provide a sensitive probe not only of the
photonic and phononic environment,
but also of memory effects such as those determined by specific initial
state preparation. At the heart of our results is an enhancement of the lack
of
photon coincidence on short time scales as a result of the exciton
superposition state. As a side-product of our findings, the
transformation from sub-Poissonian ($g^{(2)}<1$)
to super-Poissonian ($g^{(2)}>1$)
behavior on short time-scales may be of practical
interest in the design of devices which act as triggered single photon
sources. This paper has concentrated on the basic building block comprising
single nanostructures and superpositions created by the application of
one-qubit gates. Nonclassical light features in the fluorescence of coupled
QD systems will be dealt with elsewhere.

The authors acknowledge partial support from COLCIENCIAS (Colombia) projects
No.1204-05-10326, 1204-05-1148, Banco de la Rep\'ublica (Colombia)
and from the EPSRC-DTI LINK project (UK).

\newpage
\centerline{\bf Figure Captions}

\bigskip

\noindent FIG. 1. Second-order coherence function in the Markov
approximation for $\eta=2.85$, $\lambda=2.0\gamma_0$ and $\Omega=2.25\gamma_0$. (a)  $g^{(2)}$ for three different
detecting times in a Quantum Superposition (QS) initial state using the Markov
approximation. 
(b)  Time evolution of the decay rate $\gamma(t)$ and  pure
dephasing rate $\gamma_{dph}(T_e,t)$ for different temperatures.
(c)  $g^{(2)}$ including non-Markovian effects for initial QS and Ground State (GS).

\noindent FIG. 2. Second-order coherence function, including non-Markovian effects, for a Quantum Superposition (QS)
and a Statistical Mixture (SM) of states. Results shown for various Rabi frequencies
$\Omega=2\gamma_0$ (continuous line), $\Omega=2.25\gamma_0$(dotted-line), $\Omega=2.75\gamma_0$(dashed-line), 
$\Omega=6\gamma_0$(thick dot-dashed line). Here $\eta=2.85$,  $\lambda=2.0\gamma_0$ and $T\gamma_0=0.5$. 
Inset: The stationary limit for the system prepared in a QS initial state.

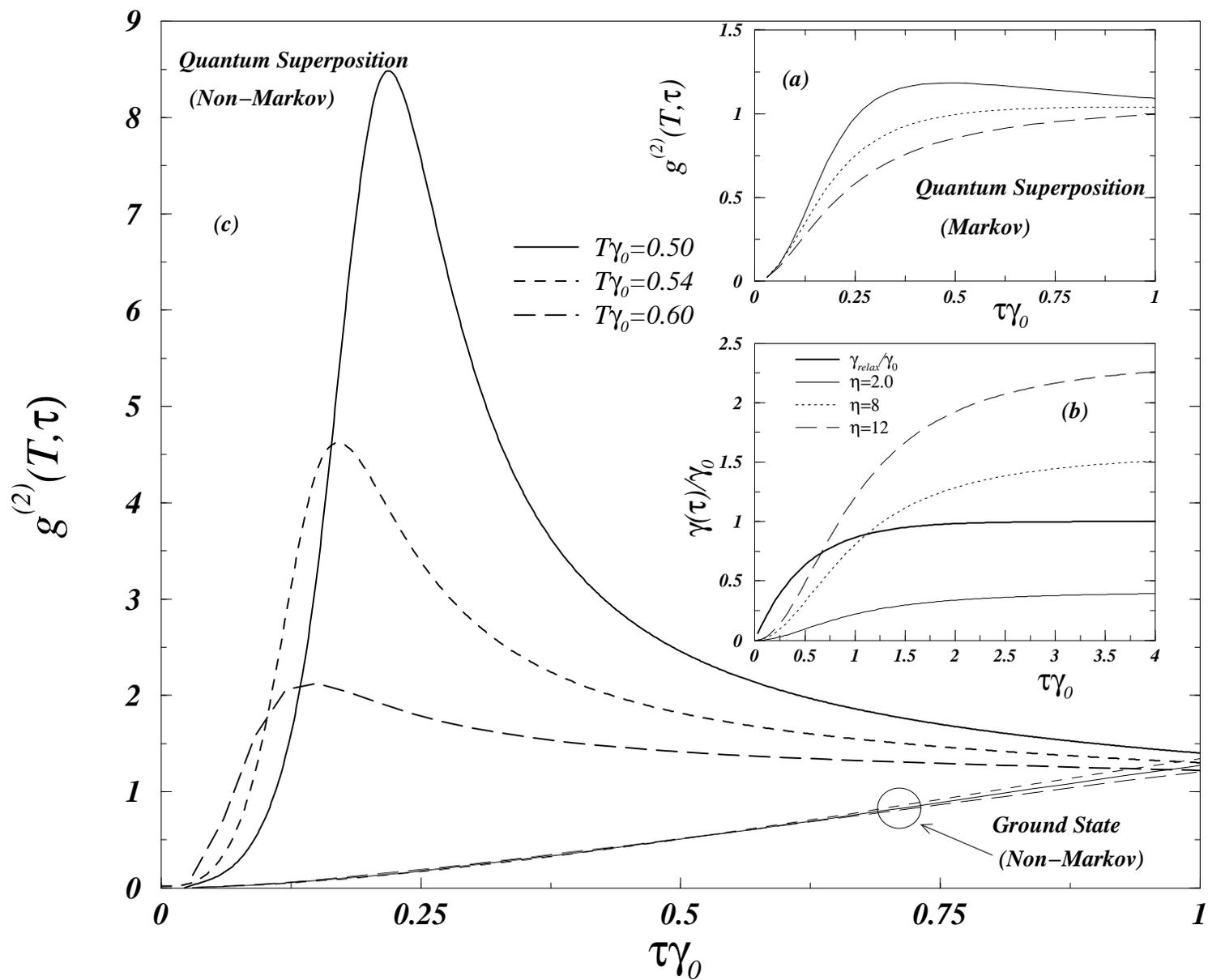
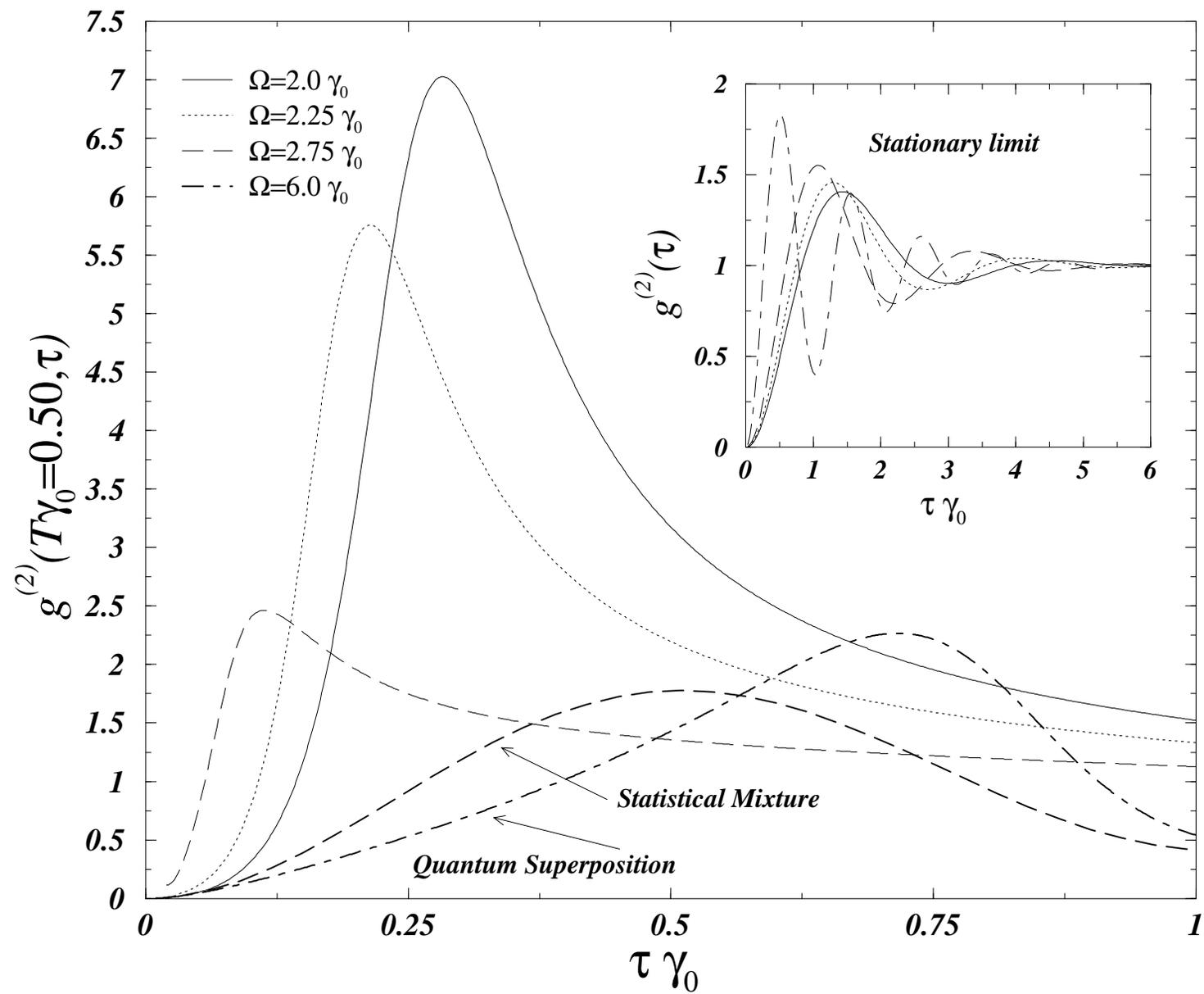
\end{document}